\documentstyle[amstex,amssymb,aps,epsf,floats]{revtex}
\tightenlines

\begin{document}
\title{Dynamics of quantum Hall stripes in double-quantum-well systems
}
\author{R. C\^{o}t\'{e}$^{1}$, H.A. Fertig$^{2}$}
\address{$^{1}$ D\'{e}partement de Physique and CERPEMA, Universit\'{e} de Sherbrooke, \\
Sherbrooke, Qu\'{e}bec, Canada J1K 2R1}
\address{$^{2}$ Department of Physics and Astronomy, University of Kentucky,%
\\
Lexington KY 40506-0055}
\date{\today }
\maketitle

\begin{abstract}
The collective modes of stripes in double layer quantum Hall
systems are computed using the time-dependent Hartree-Fock
approximation.  It is found that, when the system possesses
spontaneous interlayer coherence, there are two gapless modes,
one a phonon associated with broken translational invariance,
the other a pseudospin-wave associated with a broken
$U(1)$ symmetry.  For large layer separations the modes
disperse weakly for wavevectors perpendicular to the
stripe orientation, indicating the system becomes
akin to an array of weakly coupled one-dimensional $XY$ systems.
At higher wavevectors the collective modes develop a roton minimum
associated with a transition out of the
coherent state with further increasing layer separation.
A spin wave model of the system is developed, and it
is shown that the collective modes may be described
as those of a system with helimagnetic ordering.

PACS: 73.20.Mf,73.21.Fg,71.10.Hf
\end{abstract}

\section{Introduction and Summary of Results}

Quantum Hall systems are by now well-known to support a broad range of
condensed matter phenomena\cite{dassarma}. These include localization
physics, electron solidification (Wigner crystals), chiral Luttinger liquids
at quantum Hall edges, and Fermi liquid behavior. Recently added to this
list is highly anisotropic transport in moderate magnetic fields, in which
electrons populate several Landau levels\cite{expstr}. Such behavior was
anticipated by mean-field studies predicting unidirectional charge density
wave (CDW) ground states for electrons in high Landau levels\cite%
{shklovskii,moessner}. While more careful studies\cite{cote1} demonstrate
that within Hartree-Fock theory such ground states are unstable to the
formation of ``modulated'' stripe states which are essentially highly
anisotropic two-dimensional Wigner crystals (``stripe crystals'' \cite%
{fradkin}), quantum fluctuations may restore the translational symmetry
along the stripes\cite{fertig,yi,barci,com1}. It has been noted that the CDW
state has the symmetry of a two-dimensional smectic\cite{fradkin}, and this
analogy with liquid crystals has been exploited to yield a number of useful
results\cite{smectics}.

A seemingly unrelated set of physical phenomena occurs in double quantum
well systems (DQWS) in the quantum Hall regime. For large magnetic fields,
these systems support spontaneous interlayer coherence\cite{tlrev}, and
exhibit an associated Goldstone mode\cite{fertig2,spielman1}. A precursor of
the Josephson tunneling\cite{wen,ezawa} may have been observed\cite%
{spielman2} in these systems, contributing to recent excitement about them.

When immersed in more moderate magnetic fields, such that several Landau
levels are occupied, it becomes evident that the physics of stripes {\it and}
interlayer coherence are simultaneously relevant in the bilayer system. This
situation was recently explored\cite{brey}. For a system in which each layer
is half-filled in the $Nth$ Landau level, with some simplifying assumptions
(lower Landau levels filled and inert, Landau level mixing negligible, and
Zeeman coupling large enough that spins are polarized) it was found that:
(1) For small enough layer separation, and any value of $N$, an interlayer
coherent state of uniform density is formed. (2) Above a critical layer
separation $d_{c}(N)$, a unidirectional density modulation sets in such that
the electron density oscillates between the layers as a function of
position \cite{n=0}. 
This state is simultaneously interlayer coherent and smectic in spatial
symmetry. With increasing $d$, the regions of coherence become relatively
narrow in one direction, forming ``linear coherent regions'' (LCR's). (3)
For negligibly weak interlayer tunneling, the
coherence is spontaneous and a corresponding Goldstone mode should be
present in the excitation spectrum. When the LCR's are sufficiently narrow
compared to their separation (which occurs at large $d$ and/or large $N$),
the system may be thought of as an array of coupled one-dimensional XY
models. Different possible quantum-disordered states may exist\cite%
{brey,demler}, which in principle can be distinguished by tunneling
experiments.

In order to predict which, if any, of these quantum disorderings might
occur, it is necessary to understand the low-energy collective modes in some
detail, and to create a model that captures them which is sufficiently
simple to be susceptible to further analysis\cite{yi}. This motivates our
present study. Beyond its use in specifying a low-energy theory, the
collective modes are interesting on their own, as they are sometimes
detectible in electromagnetic absorption\cite{engel} or Raman scattering\cite%
{pinczuk} experiments, and allow one to learn about which states are
realized in real samples.

In what follows, we compute the collective mode spectrum of the electrons in
a DQWS, in a perpendicular magnetic field, in which several Landau levels
are occupied, using the time-dependent Hartree-Fock approximation (TDHFA) %
\cite{cote2}. Counting both the spin and layer index degrees of freedom, the
situations we consider involve $4N$ filled Landau levels, which are taken to
be inert (as appropriate for large cyclotron energy $\hbar \omega_c$, where $%
\omega_c = eB/m^*c$, $B$ is the magnetic field, and $m^*$ the effective mass
of electrons in the quantum wells, and large Zeeman coupling.) For
simplicity we ignore finite thickness of the wells. In addition to the $4N$
filled levels, there is a level, with Landau index $N$, which contains
precisely enough electrons so that the total filling factor $\nu$, defined
as the number of electrons per magnetic flux quantum passing through the
plane, is $4N+1$. The system considered has no interlayer bias, so that 
the highest level in each
layer will be half-filled on average.

As discussed above, if the individual layers are isolated, this is precisely
the situation in which one expects stripes to form. For finite layer
separations, it is clear that the ground state will have the electron-rich
regions in one layer align directly above the hole-rich regions in the
other. Furthermore, for a range of layer separations, the stripe edges
hybridize between the layers so that the electron occupation continuously
shifts between the layers as a function of position\cite{brey}. An example
of such a ground state is illustrated in Fig.~\ref{fig2}. 
To understand this state
and its collective modes it is convenient to adopt a pseudospin
representation, in which an electron pseudospin is ``up'' when occupying one
layer, and ``down'' when occupying the other. In this language, the stripe
ground state involves a unidirectional spatial tumbling of the spin, and may
be thought of as having helimagnetic ordering\cite{nagamiya}. As the
separation increases, the electron density tends to reside more sharply in
one well or the other, and the transition regions between layers become
narrower\cite{brey}.

In the absence of interlayer tunneling, the pseudospin density may tumble
from $+\hat{z}$ to $-\hat{z}$ by passing through the $\hat{x}$ direction,
the $\hat{y}$ direction, or any direction in between. Thus, each of the
transition regions -- the LCR's mentioned above -- supports an $XY$ degree
of freedom, which is spontaneously broken in the ground state. In addition,
this state obviously has broken translational symmetry.

Because of the two broken symmetries, there are {\it two} gapless
excitations above the stripe ground state. One is associated with the broken 
$XY$ symmetry -- alternatively thought of as a phase coherence between the
wells\cite{tlrev,fertig2} -- which disperses linearly with wavevector. The
other is a phonon-like mode which disperses quadratically for wavevectors
parallel to the stripes, and linearly perpendicular to them. This is
analogous to what is seen in incommensurate helimagnets\cite{nagamiya}.

A typical dispersion relation for the $XY$ mode is displayed in Fig.~\ref{fig5} for
several values of $d/\ell $, where $\ell =\sqrt{\hbar c/eB}$ is the magnetic
length. As may be seen, the linear mode tends to disperse more strongly
along the stripes than perpendicular to them, and in the limit of large $d$,
the latter dispersion may become quite weak. This may be understood in terms
of the exchange coupling among LCR's, which vanishes as they become
arbitrarily narrow\cite{brey}. It is possible that quantum fluctuations can
effectively wipe out the inter-LCR exchange coupling, leading to a state
analogous to a ``sliding $XY$ model''\cite{brey,ohern}.

Several other features appear in the collective mode spectrum. As is typical
of such calculations, a number of higher energy modes are present, one of
which may be interpreted as a gapped ``out-of-phase'' phonon mode, in which
the stripes in different layers oscillate against one another. In Fig.~\ref{fig5}, we
see that this mode becomes degenerate with the $XY$ mode in its dispersion
relation along the stripe direction and develops a roton minimum which
touches zero at a critical separation $d_{M}$, signaling an instability in
which modulations form along the stripes. 
The instability, however, is
first-order in nature since the modulation amplitude
changes discontinuously at the transition\cite{brey}. It
is a surprising feature of our results that a collective mode
appears to go soft so close to this transition, and this
behavior in principle allows a detection of the transition
by inelastic light scattering. 
Whether
such a bilayer stripe crystal is stable with respect to quantum fluctuations
analogous to those being discussed in the single layer case\cite%
{fertig,yi,barci,com1} is at present unclear.

This article is organized as follows. In Section II we sketch the
Hartree-Fock and TDHFA formalism, providing some details about its
application to the two-layer stripe system. Section III contains a more
detailed description of our results, and Section IV shows how the low-energy
spectrum may be understood in the pseudospin language. We conclude with a
summary and discussion in Section V.

\section{Hartree-Fock description of the CDW ground state}

We consider an unbiaised symmetric DQWS in a perpendicular magnetic field $%
{\bf B}=B{\bf z}$ at total filling factor $\nu =4N+\nu _{0}$ where $%
N=0,1,2,...$ is the Landau level index and $\nu _{0}$ is the filling factor
of the partially filled level. Each Landau level has two spin states and two
layer states specified by the index $j=R,L$. The layer states hybridize
into symmetric (S) and antisymmetric (AS)\ states in the presence of
tunneling. We assume the magnetic field to be strong enough so that the
lower $4N$ levels are completely filled with electrons and can be considered
as inert. The Zeeman energy is assumed to be much larger than the S-AS gap
and so there is no spin texture in the ground state. The electron gas is
completely spin polarised and only the layer degree of freedom need be
considered. In the ground state, the electronic charge is equally
distributed between the two wells.

In the Landau gauge where the vector potential ${\bf A}=(0,Bx,0)$, the
electron wavefunctions are given by 
\begin{equation}
\psi _{N,X,j}({\bf r})=\frac{1}{\sqrt{L_{y}}}e^{-iXy/\ell ^{2}}\varphi
_{N}\left( x-X\right) \chi _{j}(z),
\end{equation}%
where $N$ and $X$ are the Landau level and guiding center indices and $\chi
_{j}(z)$ is the envelope wave function of the lowest-energy electric subband
centered on the right or left well. $\varphi _{N}\left( x\right) $ is an
eigenfunction of the one-dimensional harmonic oscillator. Because the fully
occupied Landau levels are considered as inert, we need only consider the
partially filled Landau level. We can then drop the Landau level index from
now on.

To describe the various charge density wave (CDW) ground states, we define
the operator 
\begin{equation}
\rho _{i,j}({\bf q})=\frac{1}{N_{\varphi }}\sum_{X}e^{-iq_{x}X+iq_{x}q_{y}%
\ell ^{2}/2}\ c_{i,X}^{\dagger }c_{j,X-q_{y}\ell ^{2}},
\end{equation}%
where $i,j=R,L$ and $N_{\varphi }$ is the Landau level degeneracy. In the
ground state, $\left\langle \rho _{i,j}({\bf q})\right\rangle \neq 0$ where $%
{\bf q}$ is a reciprocal lattice vector of the CDW. By definition $%
\left\langle \rho _{R,R}({\bf q}=0)\right\rangle =\left\langle \rho _{L,L}(%
{\bf q}=0)\right\rangle =\nu _{0}/2$. In the Hartree-Fock approximation
(HFA), the ground-state energy per electron in units of $e^{2}/\kappa \ell $
in the partially filled level can be written in term of these operators as 
\begin{eqnarray}
E_{HF} &=&-\frac{2}{\nu _{0}}t%
\mathop{\rm Re}%
\left[ \left\langle \rho _{R,L}\left( {\bf q}=0\right) \right\rangle \right]
\label{ehf} \\
&&+\frac{1}{2\nu _{0}}\sum_{{\bf q}}\left[ H\left( {\bf q}\right) -X\left( 
{\bf q}\right) \right] \left[ \left| \left\langle \rho _{R,R}\left( {\bf q}%
\right) \right\rangle \right| ^{2}+\left| \left\langle \rho _{L,L}\left( 
{\bf q}\right) \right\rangle \right| ^{2}\right]  \nonumber \\
&&+\frac{1}{\nu _{0}}\sum_{{\bf q}}\widetilde{H}\left( {\bf q}\right)
\left\langle \rho _{R,R}\left( {\bf q}\right) \right\rangle \left\langle
\rho _{L,L}\left( -{\bf q}\right) \right\rangle  \nonumber \\
&&-\frac{1}{\nu _{0}}\sum_{{\bf q}}\widetilde{X}\left( {\bf q}\right)
\left\langle \rho _{R,L}\left( -{\bf q}\right) \right\rangle \left\langle
\rho _{L,R}\left( {\bf q}\right) \right\rangle .  \nonumber
\end{eqnarray}%
In this equation, $t$ is the tunneling energy in units of $e^{2}/\kappa \ell 
$, and $H\left( {\bf q}\right) ,X\left( {\bf q}\right) $ are the Hartree and
Fock {\it intra}-well interactions while $\widetilde{H}\left( {\bf q}\right) 
$ and $\widetilde{X}\left( {\bf q}\right) $ are the Hartree and Fock {\it %
inter}-well interactions. For very narrow wells where $\chi _{j}(z)$ are
highly localized, these interactions are given by 
\begin{equation}
\begin{array}{rll}
H({\bf q}) & = & \left( \frac{1}{q\ell }\right) \left[ L_{N}^{0}\left(
q^{2}\ell ^{2}/2\right) \right] ^{2}e^{-q^{2}\ell ^{2}/2}, \\ 
\widetilde{H}({\bf q}) & = & \left( \frac{1}{q\ell }\right) \left[
L_{N}^{0}\left( q^{2}\ell ^{2}/2\right) \right] ^{2}e^{-q^{2}\ell
^{2}/2}e^{-qd}, \\ 
X({\bf q}) & = & \int_{0}^{\infty }dyJ_{0}(yq\ell )e^{-y^{2}/2}\left[
L_{N}^{0}\left( y^{2}/2\right) \right] ^{2}, \\ 
\widetilde{X}({\bf q}) & = & \int_{0}^{\infty }dyJ_{0}(yq\ell
)e^{-y^{2}/2}e^{-yd/\ell }\left[ L_{N}^{0}\left( y^{2}/2\right) \right] ^{2},%
\end{array}%
\end{equation}%
where $d$ is the center-to-center separation between the wells and $%
L_{N}^{0}\left( x\right) $ are generalized Laguerre polynomials. Because of
the neutralizing positive backgrounds of ionized donors on both sides of the
DQWS, we have $H\left( 0\right) =\widetilde{H}\left( 0\right) =0$ in $E_{HF}$%
.

The order parameters $\left\{ \left\langle \rho _{i,j}\left( {\bf q}\right)
\right\rangle \right\} $ are computed by solving the HF equations of motion
for the single-particle Green's function 
\begin{equation}
G_{i,j}\left( X,X^{\prime },\tau \right) =-\left\langle Tc_{i,X}\left( \tau
\right) c_{j,X^{\prime }}^{\dagger }\left( 0\right) \right\rangle
\end{equation}
whose Fourier transform we define as 
\begin{equation}
G_{i,j}\left( {\bf q,}\tau \right) =\frac{1}{N_{\phi }}\sum_{X,X^{\prime
}}e^{-\frac{i}{2}q_{x}\left( X+X^{\prime }\right) }\delta _{X,X^{\prime
}-q_{y}l^{2}}G_{i,j}\left( X,X^{\prime },\tau \right) ,
\end{equation}%
so that $G_{i,j}\left( {\bf q,}\tau =0^{-}\right) =\left\langle \rho
_{j,i}\left( {\bf q}\right) \right\rangle .$

In the HFA, these equations of motion are given by%
\begin{gather}
\sum_{{\bf q}^{\prime \prime }}\left( 
\begin{array}{cc}
\left( i\hslash \omega _{n}+\mu \right) \delta _{{\bf q},{\bf q}^{\prime
\prime }}-U_{R,R}\left( {\bf q,q}^{\prime \prime }\right)  & t\delta _{{\bf q%
},{\bf q}^{\prime \prime }}-U_{R,L}\left( {\bf q,q}^{\prime \prime }\right) 
\\ 
t\delta _{{\bf q},{\bf q}^{\prime \prime }}-U_{L,R}\left( {\bf q,q}^{\prime
\prime }\right)  & \left( i\hslash \omega _{n}+\mu \right) \delta _{{\bf q},%
{\bf q}^{\prime \prime }}-U_{L,L}\left( {\bf q,q}^{\prime \prime }\right) 
\end{array}%
\right)  \\
\left( 
\begin{array}{cc}
G_{R,R}\left( {\bf q}^{\prime \prime },\omega _{n}\right)  & G_{R,L}\left( 
{\bf q}^{\prime \prime },\omega _{n}\right)  \\ 
G_{L,R}\left( {\bf q}^{\prime \prime },\omega _{n}\right)  & G_{L,L}\left( 
{\bf q}^{\prime \prime },\omega _{n}\right) 
\end{array}%
\right) =\left( 
\begin{array}{cc}
\hslash \delta _{{\bf q},0} & 0 \\ 
0 & \hslash \delta _{{\bf q},0}%
\end{array}%
\right)   \nonumber
\end{gather}%
where $\mu $ is the chemical potential. The effective potentials $U_{i,j}$
(in units of $e^{2}/\kappa \ell $) are defined by 
\begin{equation}
U_{R,R}\left( {\bf q,q}^{\prime }\right) =\left[ H\left( {\bf q-q}^{\prime
}\right) -X\left( {\bf q-q}^{\prime }\right) \right] \left\langle \rho
_{R,R}\left( {\bf q-q}^{\prime }\right) \right\rangle \gamma _{{\bf q},{\bf q%
}^{\prime }}+\widetilde{H}\left( {\bf q-q}^{\prime }\right) \left\langle
\rho _{L,L}\left( {\bf q-q}^{\prime }\right) \right\rangle \gamma _{{\bf q},%
{\bf q}^{\prime }},
\end{equation}%
\begin{equation}
U_{L,L}\left( {\bf q,q}^{\prime }\right) =\left[ H\left( {\bf q}-{\bf q}%
^{\prime }\right) -X\left( {\bf q}-{\bf q}^{\prime }\right) \right]
\left\langle \rho _{L,L}\left( {\bf q-q}^{\prime }\right) \right\rangle
\gamma _{{\bf q},{\bf q}^{\prime }}+\widetilde{H}\left( {\bf q}-{\bf q}%
^{\prime }\right) \left\langle \rho _{R,R}\left( {\bf q-q}^{\prime }\right)
\right\rangle \gamma _{{\bf q},{\bf q}^{\prime }},
\end{equation}%
\begin{equation}
U_{R,L}\left( {\bf q,q}^{\prime }\right) =-\widetilde{X}\left( {\bf q}-{\bf q%
}^{\prime }\right) \left\langle \rho _{L,R}\left( {\bf q-q}^{\prime }\right)
\right\rangle \gamma _{{\bf q},{\bf q}^{\prime }},
\end{equation}%
\begin{equation}
U_{L,R}\left( {\bf q,q}^{\prime }\right) =-\widetilde{X}\left( {\bf q}-{\bf q%
}^{\prime }\right) \left\langle \rho _{R,L}\left( {\bf q-q}^{\prime }\right)
\right\rangle \gamma _{{\bf q},{\bf q}^{\prime }},
\end{equation}%
where $\gamma _{{\bf q},{\bf q}^{\prime }}=e^{-i{\bf q}\times {\bf q}%
^{\prime }l_{\bot }^{2}/2}$. The procedure to solve for the $\left\langle
\rho _{i,j}\left( {\bf q}\right) \right\rangle ^{\prime }s$ is described in
detail in Ref.\cite{cote2}.

It is instructive at this point to describe the electron state in a DQWS
by using a pseudospin language where states $R$ and $L$ are mapped to up and
down spin states. For this, we define the pseudospin operators 
\begin{eqnarray}
\rho \left( {\bf q}\right) &=&\frac{1}{2}\left[ \rho _{R,R}\left( {\bf q}%
\right) +\rho _{L,L}\left( {\bf q}\right) \right] ,  \label{orderp} \\
\rho _{x}\left( {\bf q}\right) &=&\frac{1}{2}\left[ \rho _{R,L}\left( {\bf q}%
\right) +\rho _{L,R}\left( {\bf q}\right) \right] ,  \nonumber \\
\rho _{y}\left( {\bf q}\right) &=&\frac{1}{2i}\left[ \rho _{R,L}\left( {\bf q%
}\right) -\rho _{L,R}\left( {\bf q}\right) \right] ,  \nonumber \\
\rho _{z}\left( {\bf q}\right) &=&\frac{1}{2}\left[ \rho _{R,R}\left( {\bf q}%
\right) -\rho _{L,L}\left( {\bf q}\right) \right] ,  \nonumber \\
{\bf \rho }_{\bot }\left( {\bf q}\right) &=&\rho _{x}\left( {\bf q}\right) 
\widehat{{\bf x}}+\rho _{y}\left( {\bf q}\right) \widehat{{\bf y}}. 
\nonumber
\end{eqnarray}%
These operators act in the restricted Hilbert space of the partially filled
Landau level. To get a real space representation of the ordered states, we
take the Fourier transform 
\begin{equation}
\left\langle \rho _{i,j}\left( {\bf r}\right) \right\rangle =\sum_{{\bf q}%
}\left\langle \rho _{i,j}\left( {\bf q}\right) \right\rangle e^{i{\bf q}%
\cdot {\bf r}}.
\end{equation}%
A local filling factor in each well can then be defined as%
\begin{equation}
\nu _{i}\left( {\bf r}\right) =\left\langle \rho _{i,i}\left( {\bf r}\right)
\right\rangle .  \label{guiding}
\end{equation}%
For unidirectional modulations, the quantity $\nu _{i}\left( {\bf r}\right) $
will take values between $0$ and $1.$ Similarly, a Fourier transform of the
order parameters defined in Eq.~(\ref{orderp}) will define the total density 
$\left\langle \rho \left( {\bf r}\right) \right\rangle $ and spin density $%
\left\langle \rho _{\mu }\left( {\bf r}\right) \right\rangle $. Note that
these densities are related to the density of orbit centers and not to the
real density of electrons. For instance, when Landau mixing is neglected,
the real ``densities'' $n_{i,j}\left( {\bf q}\right) $ are given by%
\begin{equation}
n_{i,j}\left( {\bf q}\right) =N_{\phi }F_{N}\left( {\bf q}\right) \rho
_{i,j}\left( {\bf q}\right)
\end{equation}%
where $F_{N}\left( {\bf q}\right) $ is a form factor appropriate to Landau
level $N$ given by 
\begin{equation}
F_{N}\left( {\bf q}\right) =\exp \left( \frac{-q^{2}l^{2}}{4}\right)
L_{N}^{0}\left( \frac{q^{2}l^{2}}{2}\right) .
\end{equation}

In the pseudospin language, the HF energy can be rewritten as 
\begin{eqnarray}
E_{HF} &=&-\frac{2}{\nu _{0}}t%
\mathop{\rm Re}%
\left[ \left\langle \rho _{x}\left( 0\right) \right\rangle \right] \\
&&+\frac{1}{\nu _{0}}\sum_{{\bf q}}\Upsilon \left( {\bf q}\right)
\left\langle \rho \left( -{\bf q}\right) \right\rangle \left\langle \rho
\left( {\bf q}\right) \right\rangle  \nonumber \\
&&+\frac{1}{\nu _{0}}\sum_{{\bf q}}J_{z}\left( {\bf q}\right) \left\langle
\rho _{z}\left( -{\bf q}\right) \right\rangle \left\langle \rho _{z}\left( 
{\bf q}\right) \right\rangle  \nonumber \\
&&+\frac{1}{\nu _{0}}\sum_{{\bf q}}J_{\bot }\left( {\bf q}\right)
\left\langle {\bf \rho }_{\bot }\left( -{\bf q}\right) \right\rangle \cdot
\left\langle {\bf \rho }_{\bot }\left( {\bf q}\right) \right\rangle . 
\nonumber
\end{eqnarray}%
where the effective interactions are defined by%
\begin{eqnarray}
\Upsilon \left( {\bf q}\right) &=&H\left( {\bf q}\right) +\widetilde{H}%
\left( {\bf q}\right) -X\left( {\bf q}\right) ,  \nonumber \\
J_{z}\left( {\bf q}\right) &=&H\left( {\bf q}\right) -\widetilde{H}\left( 
{\bf q}\right) -X\left( {\bf q}\right) , \\
J_{\bot }\left( {\bf q}\right) &=&-\widetilde{X}\left( {\bf q}\right) .
\nonumber
\end{eqnarray}

We derive the dispersion relations of the collective excitations of the CDW
states in the DQWS by tracking the poles of the retarded density and
pseudospin response functions. These are obtained by analytical continuation
of the two-particle Matsubara Green's functions 
\begin{equation}
\chi _{i,j,k,l}\left( {\bf q,q}^{\prime };\tau \right) =-N_{\phi
}\left\langle T\rho _{i,j}\left( {\bf q,}\tau \right) \rho _{k,l}\left( -%
{\bf q}^{\prime },0\right) \right\rangle +N_{\phi }\left\langle \rho
_{i,j}\left( {\bf q}\right) \right\rangle \left\langle \rho _{k,l}\left( -%
{\bf q}^{\prime }\right) \right\rangle  \label{original}
\end{equation}%
which we compute in the generalized random phase approximation (GRPA). The
procedure is explained in details in Refs.\cite{cote1,cote2}. It is
convenient to work in the pseudospin language where we can define the matrix%
\begin{equation}
\chi =\left( 
\begin{array}{cccc}
\chi _{\rho ,\rho } & \chi _{\rho ,x} & \chi _{\rho ,y} & \chi _{\rho ,z} \\ 
\chi _{x,\rho } & \chi _{x,x} & \chi _{x,y} & \chi _{x,z} \\ 
\chi _{y,\rho } & \chi _{y,x} & \chi _{y,y} & \chi _{y,z} \\ 
\chi _{z,\rho } & \chi _{z,x} & \chi _{z,y} & \chi _{z,z}%
\end{array}%
\right) .
\end{equation}%
These pseudospin Green's functions are related to the original Green's
functions of eq. (\ref{original}) by the transformation%
\begin{equation}
{\chi }^{\nu ,\mu }=\frac{1}{4}\sigma _{i,j}^{\nu }\chi
_{i,j,k,l}\sigma _{k,l}^{\mu },
\end{equation}%
where $\nu ,\mu =\rho ,x,y,z.$ $\sigma _{i,j}^{\nu =x,y,z}$ are Pauli
matrices and $\sigma ^{\rho }=\left( 
\begin{array}{cc}
1 & 0 \\ 
0 & 1%
\end{array}%
\right) .$

The summation of the bubbles and ladder diagrams of the GRPA can be
expressed as an equation of motion for $\chi $ of the form%
\begin{equation}
\left( \omega +i\delta \right) \chi \left( {\bf q},{\bf q}^{\prime },\omega
\right) -\sum_{{\bf q}^{\prime \prime }}F\left( {\bf q},{\bf q}^{\prime
\prime }\right) \chi \left( {\bf q}^{\prime \prime },{\bf q}^{\prime
},\omega \right) =D\left( {\bf q},{\bf q}^{\prime }\right) .
\label{dispersion}
\end{equation}%
Here $F$ and $D$ may be written schematically as 
\begin{equation}
F=-2i\left( 
\begin{array}{cccc}
\left\langle \rho \right\rangle \sin a\left[ \Upsilon -\Upsilon ^{\prime }%
\right]  & \left\langle \rho _{x}\right\rangle \sin a\left[ J_{\bot
}-J_{\bot }^{\prime }\right]  & \left\langle \rho _{y}\right\rangle \sin a
\left[ J_{\bot }-J_{\bot }^{\prime }\right]  & \left\langle \rho
_{z}\right\rangle \sin a\left[ J_{z}-J_{z}^{\prime }\right]  \\ 
\left\langle \rho _{x}\right\rangle \sin a\left[ J_{\bot }-\Upsilon ^{\prime
}\right]  & \left\langle \rho \right\rangle \sin a\left[ \Upsilon -J_{\bot
}^{\prime }\right]  & \left\langle \rho _{z}\right\rangle \cos a\left[
J_{z}-J_{\bot }^{\prime }\right]  & -\left\langle \rho _{y}\right\rangle
\cos a\left[ J_{\bot }-J_{z}^{\prime }\right]  \\ 
\left\langle \rho _{y}\right\rangle \sin a\left[ J_{\bot }-\Upsilon ^{\prime
}\right]  & -\left\langle \rho _{z}\right\rangle \cos a\left[ J_{z}-J_{\bot
}^{\prime }\right]  & \left\langle \rho \right\rangle \sin a\left[ \Upsilon
-J_{\bot }^{\prime }\right]  & t+\left\langle \rho _{x}\right\rangle \cos a
\left[ J_{\bot }-J_{z}^{\prime }\right]  \\ 
\left\langle \rho _{z}\right\rangle \sin a\left[ J_{z}-\Upsilon ^{\prime }%
\right]  & \left\langle \rho _{y}\right\rangle \cos a\left[ J_{\bot
}-J_{\bot }^{\prime }\right]  & t-\left\langle \rho _{x}\right\rangle \cos a
\left[ J_{\bot }-J_{\bot }^{\prime }\right]  & \left\langle \rho
\right\rangle \sin a\left[ \Upsilon -J_{z}^{\prime }\right] 
\end{array}%
\right) ,  \label{f}
\end{equation}%
and%
\begin{equation}
D=\left( 
\begin{array}{cccc}
i\left\langle \rho \right\rangle \left( \sin a\right)  & i\left\langle \rho
_{x}\right\rangle \left( \sin a\right)  & i\left\langle \rho
_{y}\right\rangle \left( \sin a\right)  & i\left\langle \rho
_{z}\right\rangle \left( \sin a\right)  \\ 
i\left\langle \rho _{x}\right\rangle \left( \sin a\right)  & i\left\langle
\rho \right\rangle \left( \sin a\right)  & i\left\langle \rho
_{z}\right\rangle \cos a & -i\left\langle \rho _{y}\right\rangle \left( \cos
a\right)  \\ 
i\left\langle \rho _{y}\right\rangle \left( \sin a\right)  & -i\left\langle
\rho _{z}\right\rangle \cos a & i\left\langle \rho \right\rangle \left( \sin
a\right)  & i\left\langle \rho _{x}\right\rangle \left( \cos a\right)  \\ 
i\left\langle \rho _{z}\right\rangle \left( \sin a\right)  & i\left\langle
\rho _{y}\right\rangle \left( \cos a\right)  & -i\left\langle \rho
_{x}\right\rangle \left( \cos a\right)  & i\left\langle \rho \right\rangle
\left( \sin a\right) 
\end{array}%
\right) .  \label{d}
\end{equation}%
These matrices contain wavevector dependence that is not shown. For example,
in the $F$ matrix, the first entry is explicitly given by 
\begin{equation}
-2i\left\langle \rho \right\rangle \sin a\left[ \Upsilon -\Upsilon ^{\prime }%
\right] \rightarrow -2i\left\langle \rho \left( {\bf q}-{\bf q}^{\prime
\prime }\right) \right\rangle \sin \left[ \frac{{\bf q}\times {\bf q}%
^{\prime \prime }l_{\bot }^{2}}{2}\right] \left[ \Upsilon \left( {\bf q}-%
{\bf q}^{\prime \prime }\right) -\Upsilon \left( {\bf q}^{\prime \prime
}\right) \right] .
\end{equation}%
The other terms in the matrix, as well as those in the matrix $D$, have
analogous definitions. Finally, for the tunneling term in $F_{34}$ and $%
F_{43}$ is diagonal in the wavevector index i.e. $t\rightarrow $ $t\delta _{%
{\bf q},{\bf q}^{\prime \prime }}.$ Eq.~(\ref{dispersion}) may be solved by
diagonalizing $F$ \cite{cote2}, and the collective mode frequencies found
from its eigenvalues. From the eigenvectors of the $F$ matrix, it is also
possible to extract the motion of the guiding-center densities and of the
pseudospin in a given mode.

It is interesting to note that for 
$\left\langle \rho _{x,y}\right\rangle = 0$ and
$t=0$, the forms of $F$ and $D$ indicate that
in the equations of motion $\chi_{\rho,\rho},~
\chi_{\rho,z},~\chi_{z,\rho},~\chi_{z,z}$
completely decouple from $\chi_{xx},~\chi_{xy},~
\chi_{yx},~\chi_{yy}$.  This indicates that 
distortions of the stripes involving motion of charge 
either within the layers or between them is
completely decoupled from any ``in-plane'' $XY$
motion of the pseudospins; i.e., phonon modes and
spin-wave modes will create poles in different,
distinct response functions.  The presence of
coherence -- a non-vanishing 
$\left\langle \rho _{x}\right\rangle$ or
$\left\langle \rho _{y}\right\rangle$ --
mixes these motions, so that poles from all
the collective modes appear in all the response
functions.  This phenomenon is closely related
to ``spin-charge coupling'' that is generically
present in multicomponent quantum Hall systems \cite{dassarma},
and has important consequences for the charged excitations
in this system \cite{brey}.  For the collective modes,
we will see below that 
the low-energy interlayer charge degrees of freedom 
(distortions that change $\rho_z$) 
are in a sense conjugate to the in-plane
degrees of freedom ($\rho_{x,y}$), and distortions
of {\it both} are involved in any given collective mode.

\section{Numerical results}

\begin{figure}[tbp]
\centerline{\epsfxsize 8cm \epsffile{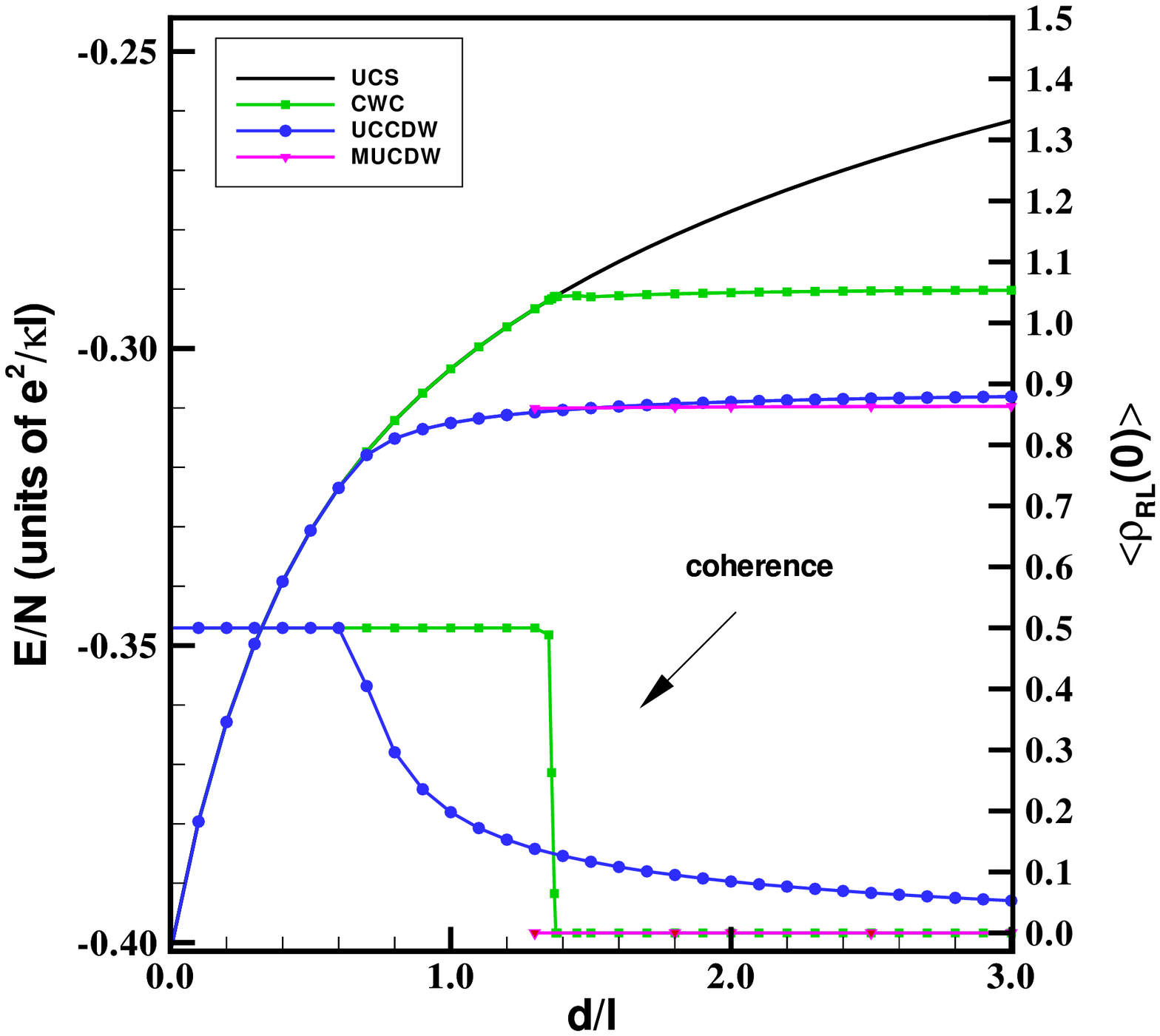}}
\caption{Hartree-Fock energy per electron and coherence as a function of the
interlayer separation in the UCS, CWC, UCCDW and MUCDW states in Landau
level $N=2$ and for $t=0$. }
\label{fig1}
\end{figure}

\begin{figure}[tbp]
\centerline{\epsfxsize 8cm \epsffile{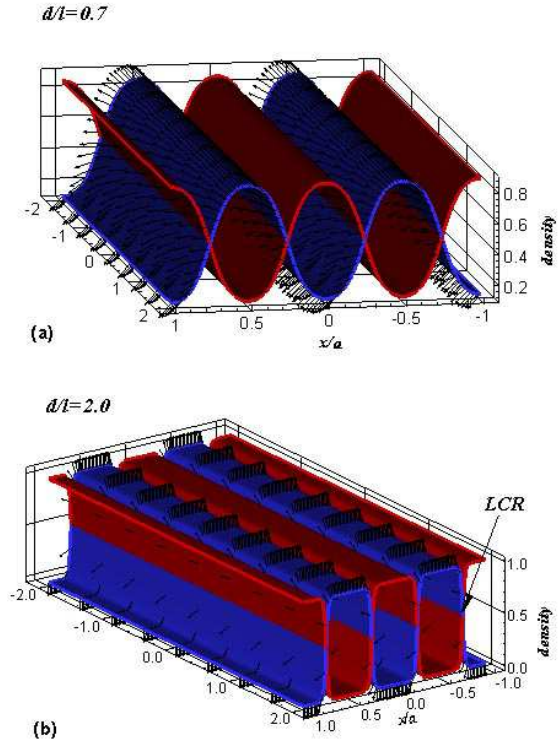}}
\caption{Real space representation of the guiding center and pseudospin
densities in the UCCDW for $N=2$ and $t=0$. (a) $d/\ell =0.7$; (b) $d/\ell
=2.0.$ The width of the linear coherent regions (LCR's) become narrower as
the interlayer separation increases.}
\label{fig2}
\end{figure}

In Ref. \cite{brey} the energies of several ordered
ground states at $\nu _{0}=1$ were computed
in the HFA. The states considered
there were a uniform coherent state (UCS), a unidirectional coherent charge
density wave (UCCDW), a modulated unidirectional charge density wave (MUCDW)
and a coherent Wigner crystal (CWC) with a square lattice. We refer the
reader to Ref. \cite{brey} for more detailed discussions of these states.
For $N>0$, it was found that the ground state of the electron gas evolves
from the UCS for $d<d_{c}\left( N\right) $ to the UCCDW at larger values of $%
d$ and finally to the MUCDW as the separation between the wells increases.
At large $d$, the Wigner crystal state is only lowest in energy for Landau
level $N=0$. (There is however a small region of $d$ for $d>d_{c}$ in $N=0$
for which the UCCDW is lowest in energy (see Ref. \cite{brey})). For all
ordered states, the lowest energy is obtained when the density pattern in
both wells are shifted with respect to one another. Moreover, coherent
states (states with non-zero value of $\left\langle \rho _{R,L}\left( {\bf q}%
\right) \right\rangle $), when they exist, have lower energy than their
incoherent counterparts. For the MUCDW it was impossible to find a coherent
version in the region where it has lower energy than the UCCDW. An
interesting result of the HF calculation is that the local coherence $%
\left\langle \rho _{R,L}\left( {\bf r}\right) \right\rangle $ is maximal
when the charge density is equally shared by both wells. For the UCCDW, this
occurs along channels called linear coherent regions (LCR). As the
separation between the wells increases, the width of the LCR's becomes very
small. Figs.~1-3 summarizes the HF results for level $N=2$. Fig.~\ref{fig1} shows the
energy of the four states defined above as a function of $d$ in Landau level 
$N=2$ and in the absence of tunneling. (We remark that all results presented
in the present paper are for $\nu _{0}=1.$ Also, in the absence of
tunneling, the phase of $\left\langle \rho _{R,L}\left( {\bf r}\right)
\right\rangle $ is arbitrarily choosen so that all spins point in the $%
\widehat{{\bf x}}$ direction in the ground state). As a measure of the
coherence of a given state, we use $\left\langle \rho _{R,L}\left( 0\right)
\right\rangle .$ This quantity takes its maximal value $1/2$ (at $\nu _{0}=1$%
) in the UCS. In Fig.~\ref{fig1}, the coherence decreases 
slowly for the UCCDW but
very rapidly for the CWC and is 
essentially zero in the MUCDW.

\begin{figure}[tbp]
\centerline{\epsfxsize 8cm \epsffile{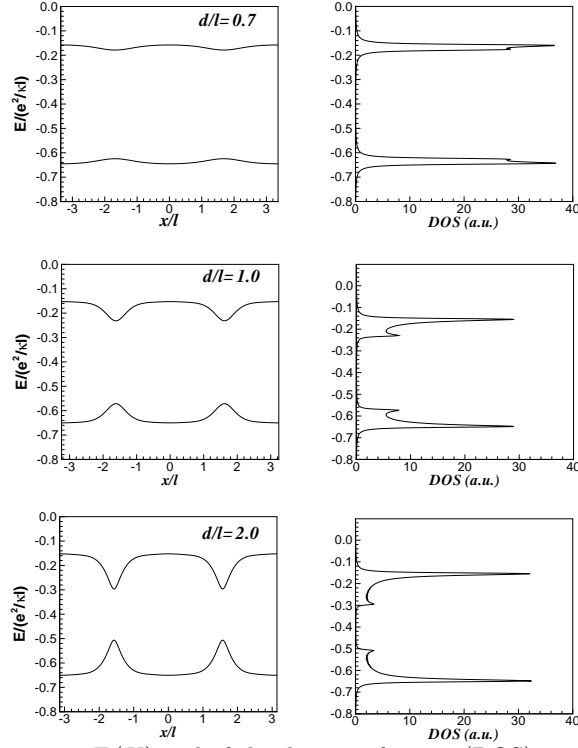}}
\caption{Evolution of the band structure $E \left(
X\right) $ and of the density of states (DOS) in the UCCDW at $N=2$ and $t=0$
as a function of the interlayer separation. The interstripes separation is
approximately $a=6.2\ell $ for these values of $d.$}
\label{fig3}
\end{figure}

In Landau level $N=2$, the UCS is lower in energy for $d/\ell <0.7$ at which
point it evolves continuously into the UCCDW. At $d/\ell \simeq 1.6$, there
is a first order transition into the MUCDW. Fig.~\ref{fig2} shows the guiding- center
density in the right and left wells as defined in Eq.~(\ref{guiding}) and
the pseudospin pattern (Eq.~(\ref{orderp})) for the UCCDW at $d/\ell =0.7$
and $d/\ell =2.0$. The formation of the LCR's is clearly visible in this
figure. At large $d$, the coherence is very small 
(see Fig.~\ref{fig1}) and the
guiding-center densities approach the stripe pattern appropriate to
decoupled layers with filling factor $1/2$. For completeness, we also show
in Fig.~\ref{fig3} the evolution of band structure\cite{brey} $E \left(
X\right) $ and of the density of states in the UCCDW at $N=2$ and for $t=0$.

We now consider the collective excitations of the UCS and UCCDW. The
dispersion is obtained by solving Eq.~(\ref{dispersion}) for the
suceptibility $\chi $. In the UCS, we can easily solve this equation to get
the response functions 
\begin{equation}
\chi =\allowbreak \left( 
\begin{array}{cccc}
0 & 0 & 0 & 0 \\ 
0 & 0 & 0 & 0 \\ 
0 & 0 & 2\frac{a}{\omega ^{2}-4ab}\alpha & i\frac{\omega }{\omega ^{2}-4ab}%
\alpha \\ 
0 & 0 & -i\frac{\omega }{\omega ^{2}-4ab}a & 2\frac{b}{\omega ^{2}-4ab}\alpha%
\end{array}%
\right) ,
\end{equation}%
where 
\begin{equation}
\alpha \equiv \left\langle \rho _{x}\left( 0\right) \right\rangle ,
\end{equation}%
and 
\begin{eqnarray}
a\left( {\bf q}\right) &=&t-\left\langle \rho _{x}\left( 0\right)
\right\rangle \left[ J_{\bot }\left( 0\right) -J_{z}\left( {\bf q}\right) %
\right], \\
b\left( {\bf q}\right) &=&t-\left\langle \rho _{x}\left( 0\right)
\right\rangle \left[ J_{\bot }\left( 0\right) -J_{\bot }\left( {\bf q}%
\right) \right] .
\end{eqnarray}%
The dispersion relation of the collective mode of the UCS is given by 
\begin{equation}
\omega =2\sqrt{a\left( {\bf q}\right) b\left( {\bf q}\right) }.
\end{equation}%
This mode is a Goldstone mode (at $t=0$) associated with the broken $XY$
symmetry of the UCS. For small wavevectors, it disperses linearly in $q$ for 
$t=0$ and $d\neq 0$. This coherence mode represents an elliptical motion of
the pseudospins around the $x$ axis. The pseudospin motion becomes more and
more confined to the $x-y$ plane as $q$ decreases. For level $N=2$ and in
the absence of tunneling, the dispersion relation of this mode softens at
interlayer separation $d/\ell \approx 0.64$. This softening occurs at a
finite value $q\ell =0.92$ of the wavevector signaling the onset of the
formation of the UCCDW state with a wavelength (separation between stripes
in a given layer) of approximately $\lambda /\ell =2\pi /0.92.$

\begin{figure}[tbp]
\centerline{\epsfxsize 8cm \epsffile{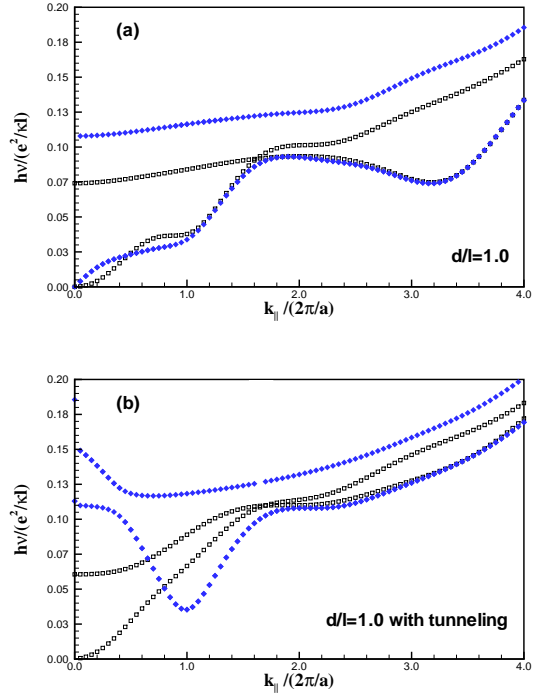}}
\caption{Dispersion relations calculated in the GRPA\ (a) without and (b)
with tunneling for the lowest-energy collective excitations in the UCCDW in
Landau level $N=2$ and for $d/\ell =1.0.$ These dispersions are for
wavevector ${\bf k}$ in the direction of the stripes ($k_{\bot }=0$). The
filled diamonds represent the phonon modes (in-phase and out-of-phase) while
the lowest two branches of the $XY$ mode are represented by the empty
squares.}
\label{fig4}
\end{figure}

\begin{figure}[tbp]
\centerline{\epsfxsize 8cm \epsffile{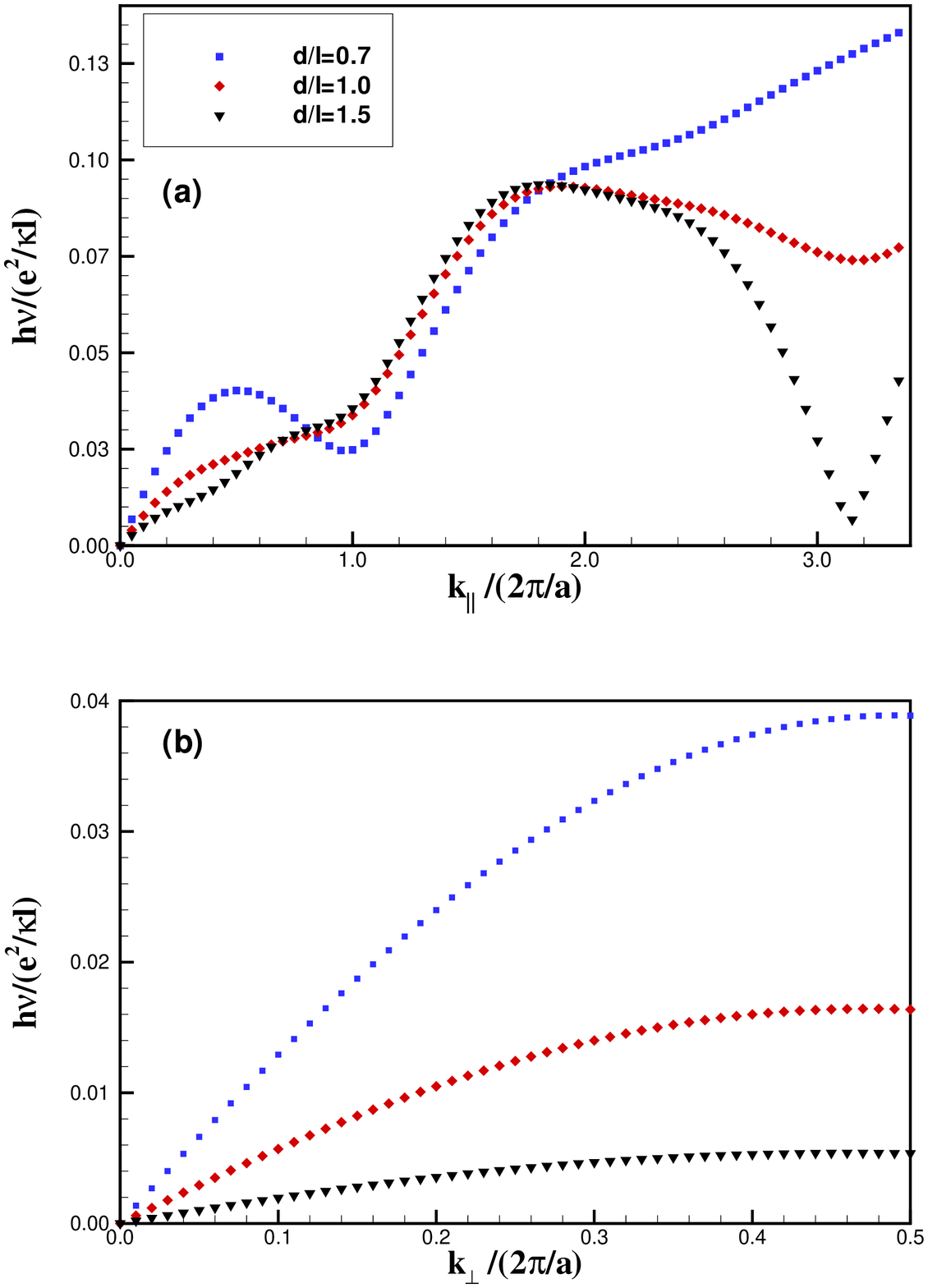}}
\caption{Dispersions of the $XY$ mode of the UCCCW with $N=2$ and $t=0$ for
wavevector (a) parallel ($k_{\bot }=0$) and (b) perpendicular ($k_{\Vert }=0$%
) to the stripes and for several values of the interlayer separation. }
\label{fig5}
\end{figure}

\begin{figure}[tbp]
\centerline{\epsfxsize 8cm \epsffile{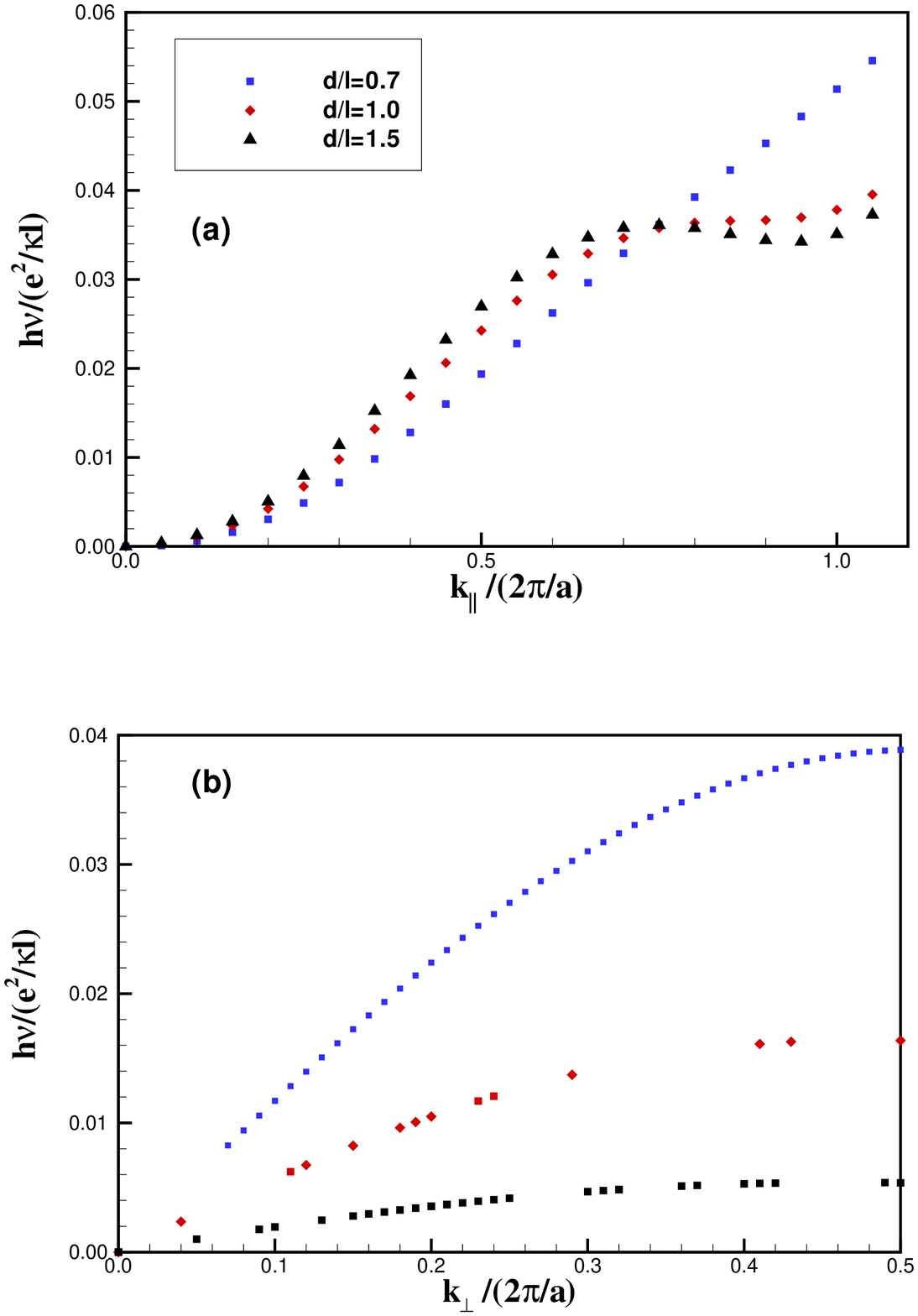}}
\caption{Dispersions of the in-phase phonon mode of the UCCCW with $N=2$ and 
$t=0$ for wavevector (a) parallel ($k_{\bot }=0$) and (b) perpendicular ($%
k_{\Vert }=0$) to the stripes and for several values of the interlayer
separation. }
\label{fig6}
\end{figure}

For our choice of phase, the pseudospins in the UCCDW rotate in the $x-z$
plane. This implies that $\left\langle \rho _{y}\left( {\bf q}\right)
\right\rangle =0.$ Moreover, in this shifted state, there is no modulation
of the {\it total} density, so that $\left\langle \rho \left( {\bf %
q}\right) \right\rangle =0$ as well. The $F$ and $D$ matrix introduced in
Eqs.~(\ref{f}) and (\ref{d}) then simplify to 
\begin{equation}
F=-2i\left( 
\begin{array}{cccc}
0 & \left\langle \rho _{x}\right\rangle \sin a\left[ J_{\bot }-J_{\bot
}^{\prime }\right]  & 0 & \left\langle \rho _{z}\right\rangle \sin a\left[
J_{z}-J_{z}^{\prime }\right]  \\ 
\left\langle \rho _{x}\right\rangle \sin a\left[ J_{\bot }-\Upsilon ^{\prime
}\right]  & 0 & \left\langle \rho _{z}\right\rangle \cos a\left[
J_{z}-J_{\bot }^{\prime }\right]  & 0 \\ 
0 & -\left\langle \rho _{z}\right\rangle \cos a\left[ J_{z}-J_{\bot
}^{\prime }\right]  & 0 & t+\left\langle \rho _{x}\right\rangle \cos a\left[
J_{\bot }-J_{z}^{\prime }\right]  \\ 
\left\langle \rho _{z}\right\rangle \sin a\left[ J_{z}-\Upsilon ^{\prime }%
\right]  & 0 & t-\left\langle \rho _{x}\right\rangle \cos a\left[ J_{\bot
}-J_{\bot }^{\prime }\right]  & 0%
\end{array}%
\right) ,  \label{f2}
\end{equation}%
and%
\begin{equation}
D=\left( 
\begin{array}{cccc}
0 & i\left\langle \rho _{x}\right\rangle \left( \sin a\right)  & 0 & 
i\left\langle \rho _{z}\right\rangle \left( \sin a\right)  \\ 
i\left\langle \rho _{x}\right\rangle \left( \sin a\right)  & 0 & 
i\left\langle \rho _{z}\right\rangle \cos a & 0 \\ 
0 & -i\left\langle \rho _{z}\right\rangle \cos a & 0 & i\left\langle \rho
_{x}\right\rangle \left( \cos a\right)  \\ 
i\left\langle \rho _{z}\right\rangle \left( \sin a\right)  & 0 & 
-i\left\langle \rho _{x}\right\rangle \left( \cos a\right)  & 0%
\end{array}%
\right) .  \label{d2}
\end{equation}%
As is clear from Eq.~(\ref{f2}), The interlayer coherence present in the
UCCDW introduces a coupling between the longitudinal and transverse response
functions. From Eqs.~(\ref{f2}) and (\ref{d2}), we see that $\chi _{\rho
,\rho }\rightarrow 0$ as $k_{\Vert }\rightarrow 0$ so that the coupling with
the density response gets very small for small wavevector parallel to the
stripes in which case the response is dominated by the pseudospin motion.
Fig.~\ref{fig4} shows the dispersion relations of the lowest four collective modes of
the UCCDW at $d/\ell =1.0$ with and without tunneling. The dispersion is
given for wavevector along the direction of the stripes with $k_{\bot }=0
$. These curves are obtained by tracking the poles in the four response
functions $\chi _{\rho ,\rho },\chi _{x,x},\chi _{y,y},\chi _{z,z}$ for
wavevector ${\bf k}$ along the desired direction in the Brillouin zone. From
the weight of given pole in each response function, we can infer the nature
of the mode. The low-energy dispersion consists of an in-phase and
out-of-phase phonon modes (empty squares in Fig.~\ref{fig4}) that both involve a
coupling between the density $\rho $ and pseudospin $\rho _{x}.$ The
in-phase phonon mode is gapless while the out-of-phase phonon is gapped.
Both phonons are gapped for $k_{\parallel}=0,~|k_{\bot}| > 0$ 
(see Fig.~\ref{fig6}(b)) in contrast with what
happens for stripes in single quantum well systems where 
the phonon frequency vanishes for all $k_{\parallel}=0$.
These behaviors are distinct because the nature of the interstripe-coupling
in the single layer and double layer systems is different in an important
way. In the single layer system, there is very little exchange interaction
between stripes. Any dispersion in the perpendicular direction comes from
direct coupling, i.e., the Hartree interaction. In the single layer case,
modulations along the stripes are present and in principle introduce a gap
in this direction. In practice, the energy cost for ``sliding'' stripes with
respect to one another is small because the modulations are weak, and is
nearly averaged out due to the long-range nature of the Coulomb potential.
Thus, in calculations such a gap is essentially immeasurable \cite{cote1}.
In the present case, the coupling between stripes is due to exchange, it is
present even in the absence of modulations along the stripes, and is not
averaged away due to the long-range nature of the interaction. The exchange
coupling is set by matrix elements between single-particle states in
different LCR's \cite{brey}; these become small in the limit of large $d$ or 
$N$ but in general are not negligible.

In addition, one may also clearly see a linearly dispersing gapless $XY$
mode in Fig.~\ref{fig4} which, as in the UCS, represents a motion of the spins in the 
$y-z$ plane. In the UCCDW, the dispersion relation of this mode is folded
into the first Brillouin zone (as is the case for the phonon modes as well).
In Fig.~\ref{fig4}, we show two branches of this $XY$ mode represented by the filled
diamonds. Fig.~\ref{fig4}(b) shows how tunneling affects these dispersion relations.
As expected, the phonon modes are not dramatically affected by switching on
the tunneling while the phase ($XY$) mode becomes gapped.

The dispersion relations of the phase and in-phase phonon modes are plotted
in Figs.~\ref{fig5} and \ref{fig6} for directions parallel and perpendicular 
to the stripes and
for several values of $d/l.$ For the phase mode, the dispersion is linear in
both directions but weaker in the perpendicular direction. Comparing Figs.~\ref{fig6}(a) 
and (b), we see that the phonon dispersion is quadratic along the
stripes and linear for direction perpendicular to the stripes. As $d$
increases, the phonon and phase mode dispersions in the perpendicular
direction become very weak and eventually their gaps vanish in the limit of
very large $d$. The suppression of these gaps reflects the shrinking of the
exchange coupling discussed above, and indicates that the system is
essentially an array of weakly coupled one-dimensional systems in this
limit. For the phonon mode, our results are consistent with the calculated
dispersion for the phonon mode of stripes in single quantum well \cite{cote1}.
 The qualitative behavior of the gapless modes at low energies may be
understood in terms of a spin-wave model which will be developed in the next
section.

In Fig.~\ref{fig4}(a), the out-of-phase phonon mode is seen to become degenerate with
the $XY$ mode at large values of $k_{\Vert }.$ Both modes soften at
approximately $q_{\Vert }/\left( \frac{2\pi }{a}\right) =3.1$ when $d$
increases and at $d/\ell =1.6$ ($a$ is the separation between the stripes in
a given layer) they become unstable. The period of modulation along the
stripes implied by this instability is consistent with the formation of a
MUCDW or highly anisotropic Wigner crystal with one electron per unit cell
in each well. 
The softening apparently accompanies a first order transition
into the MUCDW, since both the UCCDW and MUCDW exist as solutions
to the HFA both above and below the critical separation, and cross
in energy very close to it.
Note that for stripes in a single
quantum well system, within HF theory the unidirectional CDW state is always
unstable with respect to the formation of an anisotropic Wigner crystal.
Here, the instability only occurs at large enough values of $d$. In the
MUCDW, the coherence is quickly lost 
with increasing $d$ as can be seen on Fig.~\ref{fig1}.

\section{Spin Wave Analysis}

As mentioned previously, the symmetry of the ground state and the low-energy
excitations are formally quite similar to those of a non-collinear
ferromagnet, with helimagnet ordering\cite{nagamiya}. In this section, we
demonstrate that such a model can be constructed that captures the
low-energy behavior, and explicitly demonstrates the origin of the two
low-energy modes. Our analysis uses the pseudospin analogy introduced above
but now with the real electron density difference between wells which we
denote by 
\begin{equation}
S_{z}\left( {\bf r}\right) ={\frac{1}{2}}\left[ \psi _{R}^{\dag }\left( {\bf %
r}\right) \psi _{R}\left( {\bf r}\right) -\psi _{L}^{\dag }\left( {\bf r}%
\right) \psi _{L}\left( {\bf r}\right) \right] .  \label{Sz}
\end{equation}%
With the natural definitions for spin raising and lowering operators, 
\begin{equation}
S_{+}\left( {\bf r}\right) =S_{-}\left( {\bf r}\right) ^{\dag }=\psi
_{R}^{\dag }\left( {\bf r}\right) \psi _{L}\left( {\bf r}\right)   \label{S+}
\end{equation}%
we then have in-plane spin components $S_{x}={\frac{1}{2}}\left[ S_{+}+S_{-}%
\right] $, $S_{y}={\frac{1}{{2i}}}\left[ S_{+}-S_{-}\right] $. These spin
operators obey the usual commutation relations 
\begin{equation}
\left[ S_{i}\left( {\bf r}\right) ,S_{j}\left( {\bf r}^{\prime }\right) %
\right] =\sum_{k }\varepsilon _{ijk}S_{k}\left( {\bf r}\right) \delta
\left( {\bf r-r}^{\prime }\right) 
\end{equation}%
where $i,j,k=x,y,z$ and $\varepsilon _{ijk}$ is the antisymmetric tensor.

These spin operators are obviously related to the spin operators defined in
Section II. The connection is most easily seen when the spin commutation
relations are Fourier transformed, to give 
\begin{equation}
\lbrack S_{x}\left( {\bf q}\right) ,S_{y}\left( {\bf q}^{\prime }\right)
]=iS_{z}\left( {\bf q+q}^{\prime }\right) .  \label{usual}
\end{equation}%
This should be compared to the guiding center density and spin density
operators [Eqs.~(\ref{orderp})] which obey the algebra

\begin{eqnarray}
N_{\phi }\left[ \rho \left( {\bf q}\right) ,\rho \left( {\bf q}^{\prime
}\right) \right] &=&-i\sin \left( {\bf q}\times {\bf q}^{\prime
}l^{2}/2\right) \rho \left( {\bf q+q}^{\prime }\right)  \nonumber \\
N_{\phi }\left[ \rho \left( {\bf q}\right) ,\rho _{i}\left( {\bf q}^{\prime
}\right) \right] &=&-i\sin \left( {\bf q}\times {\bf q}^{\prime
}l^{2}/2\right) \rho _{i}\left( {\bf q+q}^{\prime }\right) ,  \nonumber \\
N_{\phi }\left[ \rho _{i}\left( {\bf q}\right) ,\rho _{j}\left( {\bf q}%
^{\prime }\right) \right] &=&i\varepsilon _{ijk}\cos \left( {\bf q}\times 
{\bf q}^{\prime }l^{2}/2\right) \rho _{k}\left( {\bf q+q}^{\prime }\right)
.\qquad i\neq j  \label{commutateur}
\end{eqnarray}%
It is clear that in the limit of small ${\bf q}$ and ${\bf q}^{\prime }$ the
density and spin density operators decouple. Moreover, if we make the
identification ${S}_{i}\left( {\bf q}\right) \sim $ $N_{\phi }\rho
_{i}\left( {\bf q}\right) $ one can see the direct connection between the
microscopic operators and the effective ones used in this section in the
long-wavelength limit \cite{comcom}.

If we are interested in just the low-energy, long-wavelength physics, our
two-layer system should be describable in terms of these operators\cite%
{tlrev}. The most general quadratic Hamiltonian we can write down for the
system that is consistent with the $U\left( 1\right) $ symmetry in the
absence of tunneling takes the form 
\begin{equation}
H=\int d{\bf r}\int d{\bf r}^{\prime }\left\{ K_{\parallel }\left( {\bf r-r}%
^{\prime }\right) \left[ S_{x}\left( {\bf r}\right) S_{x}\left( {\bf r}%
^{\prime }\right) +S_{y}\left( {\bf r}\right) S_{y}\left( {\bf r}^{\prime
}\right) \right] +K_{\perp }\left( {\bf r-r}^{\prime }\right) S_{z}\left( 
{\bf r}\right) S_{z}\left( {\bf r}^{\prime }\right) \right\} .
\label{H_spin}
\end{equation}%
The functions $K_{\parallel }$ and $K_{\perp }$ are assumed to have a form
that will induce a spin density wave, i.e., stripes, in the ground state.
For example, they could take the (Fourier transformed) form $K_{\parallel
}\left( q\right) =\rho _{s}q^{2}$, $K_{\perp }\left( q\right) =\kappa \left(
-q^{2}+q^{4}\xi ^{2}\right) $ where $q$ is the wavevector, $\rho _{s}$ a
spin stiffness, and $\kappa $, $\xi ^{2}$ are positive constants. One can
see for $\kappa >\rho _{s}$ that a uniform spin state will be unstable to a
state in which the spins tumble spatially; {\it i.e.}, helimagnetic
ordering. The precise form of the ground state is difficult to find, even if
the spin operators are treated classically; however, qualitatively we know
they will have a form similar to the stripe states in our Hartree-Fock
analysis. In any case, the results below do not depend on any specific
choice of $K_{\parallel }$ or $K_{\perp }$, only on the requirement that
there is helimagnetic ordering in the ground state.

As is common in a spin-wave analysis\cite{nagamiya}, we begin by treating
the spins classically. Imposing the constraint $S_{x}\left( {\bf r}\right)
^{2}+S_{y}\left( {\bf r}\right) ^{2}+S_{2}\left( {\bf r}\right) ^{2}=1$ with
a Lagrange multiplier $\lambda \left( {\bf r}\right) $, Eq.~(\ref{H_spin})
may be minimized to obtain the three equations 
\begin{eqnarray}
\int d{\bf r}^{\prime }K_{\parallel }\left( {\bf r-r}^{\prime }\right)
S_{x,y}\left( {\bf r}^{\prime }\right) &=&\lambda \left( {\bf r}\right)
S_{x,y}\left( {\bf r}\right)  \nonumber \\
\int d{\bf r}^{\prime }K_{\perp }\left( {\bf r-r}^{\prime }\right)
S_{z}\left( {\bf r}^{\prime }\right) &=&\lambda \left( {\bf r}\right)
S_{z}\left( {\bf r}\right) .  \label{spinstate}
\end{eqnarray}%
Eqs.~(\ref{spinstate}) together with the constraint equation specify the
(classical) ground state. We assume the solutions to these equations may be
written in the form 
\begin{eqnarray}
S_{x}\left( {\bf r}\right) &=&\cos \theta \left( x\right) \equiv c\left(
x\right)  \nonumber \\
S_{y}\left( {\bf r}\right) &=&0  \nonumber \\
S_{z}\left( {\bf r}\right) &=&\sin \theta \left( x\right) \equiv s\left(
x\right) .  \label{sc}
\end{eqnarray}%
Given the symmetries of $H$, it is clear that equal energy, inequivalent
states can be generated by rotating ${\bf S}$ in the $x-y$ (spin) plane, by
translation [$\theta \left( x\right) \rightarrow \theta \left( x-u\right) $
for $u$ a constant], or by rotation [$\theta \left( x\right) \rightarrow
\theta \left( \widehat{{\bf Q}}\cdot {\bf r}\right) ,$with$\left| \widehat{%
{\bf Q}}\right| =1$]. These properties are responsible for the presence of
the two gapless modes and their dispersions.

The spin wave spectrum around this ground state is conveniently found by
working in a rotated spin basis, such that in the ground state all the spins
are aligned along the $\hat{z}$ axis\cite{nagamiya}. We thus define new spin
operators 
\begin{equation}
\left( 
\begin{array}{c}
S_{x}^{\prime }\left( {\bf r}\right) \\ 
S_{y}^{\prime }\left( {\bf r}\right) \\ 
S_{z}^{\prime }\left( {\bf r}\right)%
\end{array}%
\right) =\left( 
\begin{array}{ccc}
s\left( x\right) & 0 & -c\left( x\right) \\ 
0 & 1 & 0 \\ 
c\left( x\right) & 0 & s\left( x\right)%
\end{array}%
\right) \left( 
\begin{array}{c}
S_{x}\left( {\bf r}\right) \\ 
S_{y}\left( {\bf r}\right) \\ 
S_{z}\left( {\bf r}\right)%
\end{array}%
\right) .  \label{spin_rot}
\end{equation}%
Expanding $S_{z}^{\prime }\equiv \sqrt{1-S_{x}^{\prime ^{2}}-S_{y}^{\prime
^{2}}}\approx 1-{\frac{1}{2}}\left[ S_{x}^{\prime ^{2}}+S_{y}^{\prime ^{2}}%
\right] $ and making use of Eqs.~(\ref{H_spin}), (\ref{spinstate}), and (\ref%
{spin_rot}), to quadratic order in ${\bf S}^{\prime }$ after some algebra
the Hamiltonian may be written as 
\begin{equation}
H-E_{0}=\int d{\bf r}\int d{\bf r}^{\prime }\left[ K^{xx}\left( {\bf r,r}%
^{\prime }\right) S_{x}^{\prime }\left( {\bf r}\right) S_{x}^{\prime }\left( 
{\bf r}^{\prime }\right) +K^{yy}\left( {\bf r-r}^{\prime }\right)
S_{y}^{\prime }\left( {\bf r}\right) S_{y}^{\prime }\left( {\bf r}^{\prime
}\right) \right]  \label{H_rot}
\end{equation}%
where 
\begin{equation}
K^{xx}\left( {\bf r,r}^{\prime }\right) =s\left( x\right) K_{\parallel
}\left( {\bf r-r}^{\prime }\right) s\left( x^{\prime }\right) +c\left(
x\right) K_{\perp }\left( {\bf r-r}^{\prime }\right) c\left( x^{\prime
}\right) -\lambda \left( {\bf r}\right) \delta \left( {\bf r-r}^{\prime
}\right) ,
\end{equation}%
\begin{equation}
K^{yy}\left( {\bf r-r}^{\prime }\right) =K_{\parallel }\left( {\bf r-r}%
^{\prime }\right) -\lambda \left( {\bf r}\right) \delta \left( {\bf r-r}%
^{\prime }\right) ,
\end{equation}%
and the ground state energy is 
\begin{equation}
E_{0}=\int d{\bf r}\int d{\bf r}^{\prime }\left[ c\left( x\right)
K_{\parallel }\left( {\bf r-r}^{\prime }\right) c\left( x^{\prime }\right)
+s\left( x\right) K_{\perp }\left( {\bf r-r}^{\prime }\right) s\left(
x^{\prime }\right) \right] .
\end{equation}

In the classical ground state, $S_{z}^{\prime }=1$. The spin wave
approximation amounts to approximating the spin commutation relation between 
$S_{x}^{\prime }\left( {\bf r}\right) $ and $S_{y}^{\prime }\left( {\bf r}%
^{\prime }\right) $ by 
\begin{equation}
\lbrack S_{x}^{\prime }\left( {\bf r}\right) ,S_{y}^{\prime }\left( {\bf r}%
^{\prime }\right) ]=i\delta \left( {\bf r-r}^{\prime }\right) S_{z}^{\prime
}\left( {\bf r}\right) \approx i\delta \left( {\bf r-r}^{\prime }\right) .
\end{equation}%
Eq.~(\ref{H_rot}) is particularly easy to work with, because these
commutation relations allow us to think of $S_{x}^{\prime }$ as a
generalized ``position'', and $S_{y}^{\prime }$ as a ``momentum'', and in
the Hamiltonian they are decoupled.

\begin{figure}[tbp]
\centerline{\epsfxsize 8cm \epsffile{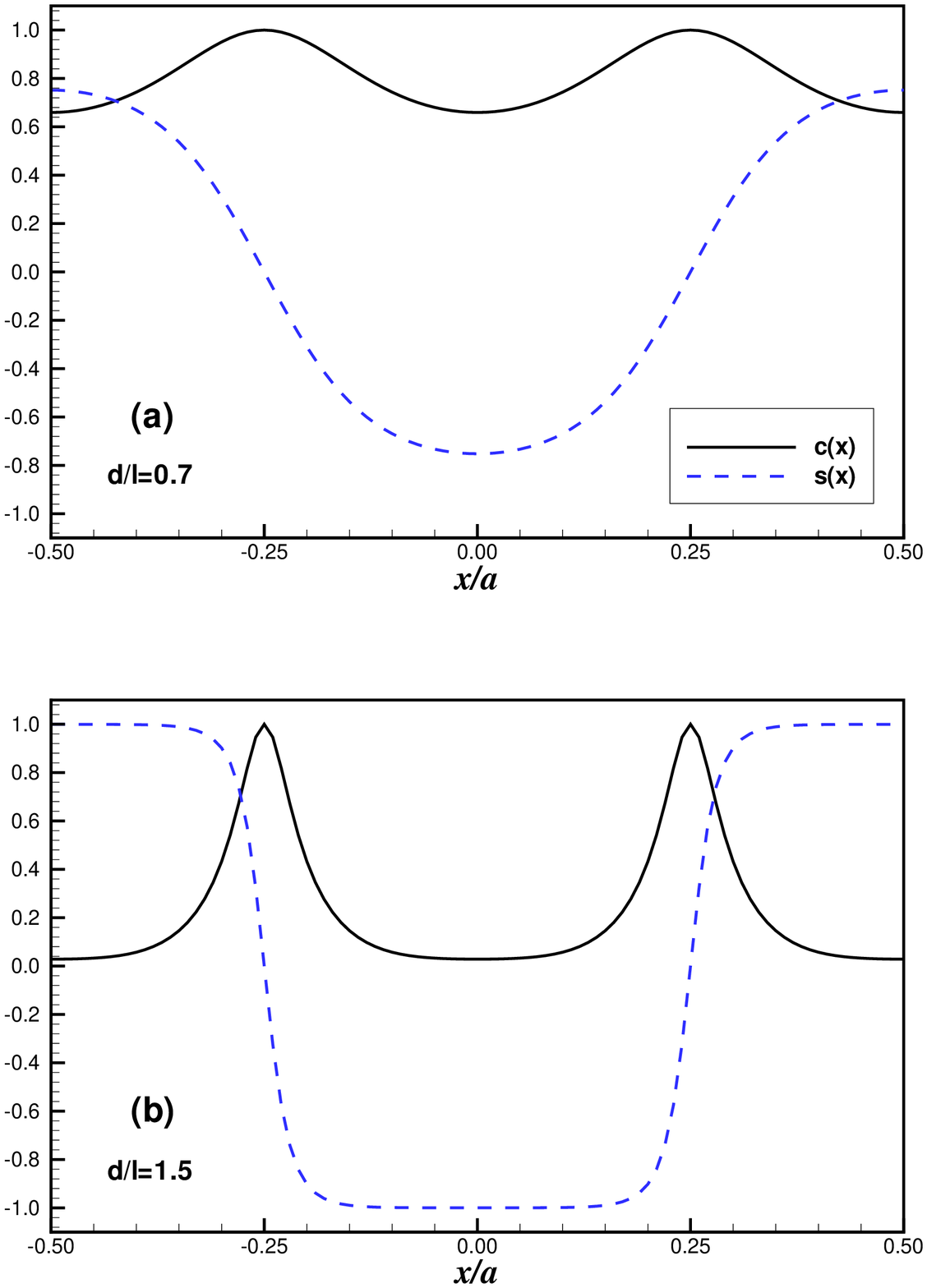}}
\caption{Plots of the functions $c(x)$ and $s(x)$ defined in Eq.~(\ref{sc})
for the UCCDW in $N=2$ and for $t=0$. (a) $d/\ell =0.7$; (b) $d/\ell =1.5$.}
\label{fig7}
\end{figure}

An exact computation of the normal modes of Eq.~(\ref{H_rot}) is quite
difficult; however, we can understand the basic properties of the spectrum
through the symmetries of the Hamiltonian and the ground state. Fig.~\ref{fig7}
illustrates the shapes of $s\left( x\right) $ and $c\left( x\right) $ in the
stripe state for two values of $d/\ell $. 
Taking $a$ to be the distance between
stripe centers in a single layer
({\it i.e}., the width of a full unit cell), it is
interesting and important to notice that $H$ is invariant under the
operation ${\bf S}^{\prime }\left( {\bf r}\right) \rightarrow {\bf S}%
^{\prime }\left( {\bf r}+\frac{a}{2}\widehat{{\bf x}}\right) $; i.e., the
primitive unit cell is half the size one expects naively, because $s\left(
x\right) $ enters the Hamiltonian quadratically, and is invariant under $%
s\left( x\right) \rightarrow -s\left( x+a/2\right) $. The discrete
translational invariance tells us that the normal modes should be expanded
in Bloch functions: 
\begin{eqnarray}
u_{n}\left( {\bf q}\right) =\int d^{2}r &h_{n,q_{x}}^{\ast }\left( x\right)
&e^{-i{\bf q}\cdot {\bf r}}S_{x}^{\prime }\left( {\bf r}\right)  \\
p_{n}\left( {\bf q}\right) =\int d^{2}r &h_{n,q_{x}}\left( x\right) &e^{i%
{\bf q}\cdot {\bf r}}S_{y}^{\prime }\left( {\bf r}\right)  \\
h_{n,q_{x}}\left( x+a/2\right)  &=&h_{n,q_{x}}\left( x\right)  \\
\int_{0}^{a/2}dx h_{n,q_{x}}^{\ast }\left( x\right) h_{m,q_{x}}\left( x\right) 
&=&\delta _{mn}
\end{eqnarray}%
and $-2\pi /a\leq q_{x}<2\pi /a$ defines 
the effective Brillouin zone. The functions $%
h_{n,q_{x}}$ may be chosen so that $H-E_{0}$ takes the form 
\begin{equation}
H-E_{0}=\Omega \sum_{n}\int {\frac{{d}{\bf q}}{{(2\pi )^{2}}}}\left[ {\frac{1%
}{2}}k_{n}\left( {\bf q}\right) u_{n}\left( -{\bf q}\right) u_{n}\left( {\bf %
q}\right) +\frac{1}{2m_{n}\left( {\bf q}\right) }p_{n}\left( -{\bf q}\right)
p_{n}\left( {\bf q}\right) \right] ,  \label{Hpm}
\end{equation}%
where $\Omega $ is the system area, and 
\begin{eqnarray}
{\frac{1}{2}}k_{n}\left( {\bf q}\right) ={\frac{1}{{\Omega }}}\int d{\bf r}%
\int d{\bf r}^{\prime }h_{n,q_{x}}^{\ast }\left( x\right) e^{-i{\bf q}\cdot 
{\bf r}} &K^{xx}\left( {\bf r,r}^{\prime }\right) &e^{i{\bf q}\cdot {\bf r}%
^{\prime }}h_{n,q_{x}}\left( x^{\prime }\right)   \nonumber \\
\frac{1}{2m_{n}\left( {\bf q}\right) }={\frac{1}{{\Omega }}}\int d{\bf r}%
\int d{\bf r}^{\prime }h_{n,q_{x}}\left( x\right) e^{i{\bf q}\cdot {\bf r}}
&K^{yy}\left( {\bf r-r}^{\prime }\right) &e^{-i{\bf q}\cdot {\bf r}^{\prime
}}h_{n,q_{x}}^{\ast }\left( x^{\prime }\right) .  \label{km}
\end{eqnarray}

From the form of Eq.~(\ref{Hpm}), it is clear that the excitation
frequencies of the system are $\omega _{n}^{2}\left( {\bf q}\right)
=k_{n}\left( {\bf q}\right) /m_{n}\left( {\bf q}\right) .$ We thus see that
there will be gapless (zero) modes whenever $k_{n}\left( {\bf q}\right) $ or 
$1/m_{n}\left( {\bf q}\right) $ vanishes. This occurs if there are choices
for the Bloch functions $h_{n,q_{x}}\left( x\right) $ which satisfy either 
\begin{equation}
\int d{\bf r}^{\prime }K^{xx}\left( {\bf r,r}^{\prime }\right) e^{i{\bf q}%
\cdot {\bf r}^{\prime }}h_{n,q_{x}}\left( x^{\prime }\right) =0  \label{van1}
\end{equation}%
or 
\begin{equation}
\int d{\bf r}^{\prime }K^{yy}\left( {\bf r-r}^{\prime }\right) e^{-i{\bf q}%
\cdot {\bf r}^{\prime }}h_{n,q_{x}}^{\ast }\left( x^{\prime }\right) =0.
\label{van2}
\end{equation}%
Using the symmetries of the ground state, one may find two choices of $%
h_{n,q_{x}}\left( x\right) $ satisfying Eq.~(\ref{van1}) or (\ref{van2}).
For 
\begin{equation}
h_{n=0,q_{x}=0}^{\ast }\left( x\right) =c\left( x\right) \equiv \cos \theta
\left( x\right)   \label{mode1}
\end{equation}%
it is easily shown that Eq.~(\ref{van2}) is satisfied. This mode represents
a uniform rotation of the ground state spin pattern ${\bf S}\left( {\bf r}%
\right) $ in the spin $x-y$ plane; {\it i.e.}, it is associated with the
spontaneous phase coherence in the ground state. Making use of Eqs.~(\ref%
{spinstate}), one may also show that Eq.~(\ref{van1}) is satisfied for 
\begin{equation}
h_{n=0,q_{x}=Q/2}\left( x\right) ={\frac{{d\theta }}{{dx}}}e^{-i{\bf Q}\cdot 
{\bf r}/2}  \label{mode2}
\end{equation}%
for ${\bf Q}={\frac{{4\pi }}{a}}\widehat{x}$. This second zero mode arises
due to the translational invariance in $H$ and is a phonon mode. It is
interesting to note that the two zero modes are found in different parts of
the Brillouin zone, the phase mode dispersing from the zone center, the
phonon from the zone edge of the effective Brillouin zone. In our
numerical calculations we found both modes dispersing from the zone center.
The reason for this is that our numerical technique obliges us to work with
the naive primitive unit cell, with a resulting Brillouin zone half the size
of the one we use in this section, so that the phonon mode is folded back to
the zone center. It is important to note that since two zero modes occur at
different wavevectors, they do not mix together and complicate the
dispersion $\omega _{n}\left( {\bf q}\right) $. This point was missed in
Ref. \cite{brey}, where it was supposed that such mixing would lead to
only a single gapless mode with appreciable oscillator strength in most
response functions. The presence of two gapless modes dispersing from
different points in the Brillouin zone is precisely what one finds for the
spin wave spectrum of an incommensurate helimagnet \cite{nagamiya}.

We are left with determining how $\omega _{0}\left( {\bf q}\right) $
disperses from the two zero modes. In the case of the phase mode, for which $%
1/m_{0}\left( {\bf q}\right) $ vanishes at ${\bf q}=0$, we may approximate $%
k_{n=0,q_{x}}\approx k_{n=0,q_{x}=0}\equiv k_{0}$ near ${\bf q}=0$, and
expand $1/m_{0}\left( {\bf q}\right) $ in small powers of ${\bf q}$ to find 
\begin{equation}
\omega _{0}^{2}\left( {\bf q}\right) \approx k_{0}\left[ \rho _{s}^{\perp
}q_{x}^{2}+\rho _{s}^{\parallel }q_{y}^{2}\right]  \label{m1_disp}
\end{equation}%
where 
\begin{eqnarray}
\rho _{s}^{\perp } &=&-{\frac{1}{\Omega }}\int d{\bf r}\int d{\bf r}^{\prime
}c\left( x\right) K_{\parallel }\left( {\bf r-r}^{\prime }\right) \left(
x-x^{\prime }\right) ^{2}c\left( x^{\prime }\right) , \\
\rho _{s}^{\parallel } &=&-{\frac{1}{\Omega }}\int d{\bf r}\int d{\bf r}%
^{\prime }c\left( x\right) K_{\parallel }\left( {\bf r-r}^{\prime }\right)
\left( y-y^{\prime }\right) ^{2}c\left( x^{\prime }\right) .
\end{eqnarray}%
Thus, the phase mode disperses linearly from ${\bf q}=0$. Notice that if $%
c\left( x\right) $ is only very different from zero in narrow regions, as
occurs for large layer separations (see Fig.~\ref{fig7}), then $c\left( x\right)
K_{\parallel }\left( {\bf r-r}^{\prime }\right) c\left( x^{\prime }\right) $
will be small unless $x$ and $x^{\prime }$ are in the same LCR. Due to the $%
\left( x-x^{\prime }\right) ^{2}$ factor in $\rho _{s}^{\perp }$, $\rho
_{s}^{\perp }\ll \rho _{s}^{\parallel }$ in this limit; {\it i.e.}, the
dispersion of the phase mode perpendicular to the stripes becomes relatively
weak. This is precisely the behavior observed in our numerical
calculations of the collective modes.

For the phonon mode, it is $k_{n=0,{\bf q}}$ which vanishes as ${\bf q}%
\rightarrow {\bf Q}/2$. Writing $\delta {\bf q}={\bf q}-{\bf Q}/2$, it is
not difficult to see how $k_{n=0,q_{x}}$ must behave for small $|\delta {\bf %
q}|$, once one recognizes that the ``position'' field $u_{n=0,{\bf q}={\bf Q}%
/2+\delta {\bf q}}$, when Fourier transformed back to real space represent a
displacement perpendicular to the stripes. In this case, the stiffness
must have the standard smectic form\cite{cote1} $k_{n=0,{\bf q}={\bf Q}%
/2+\delta {\bf q}}=\kappa _{\perp }\delta q_{x}^{2}+\kappa _{b}\delta
q_{y}^{4}$ where $\widehat{{\bf y}}$ is the direction parallel to the
stripes, and the absence of a $\delta q_{y}^{2}$ term is a direct result of
the rotational symmetry of the Hamiltonian. $\kappa _{b}$ is a bending
modulus for the stripes and represents an energy cost for introducing a
curvature along them. Writing $m_{0,{\bf q}={\bf Q}/2+\delta \vec{q}}\approx
m_{0}$, we find 
\begin{equation}
\omega _{0}^{2}({\bf Q}/2+\delta {\bf q})\approx \frac{1}{m_{0}}\left[
\kappa _{\perp }\delta q_{x}^{2}+\kappa _{b}\delta q_{y}^{4}\right] .
\label{m2_disp}
\end{equation}%
Thus, we see the phonon mode disperses linearly with $\left| \delta {\bf q}%
\right| $, except along the direction parallel to the stripes, for which it
disperses quadratically. A careful examination of the phonon mode in our
numerical results confirms this behavior.

In closing this section, we note that an observation of the phonon mode
dispersion would yield a direct confirmation of stripe ordering in this
system: the quadratic dispersion along the stripes is indicative of
spontaneous smectic ordering.

\section{Conclusion}

In a double layer system, the uniform coherent state (UCS) is unstable with
respect to the formation of a unidirectional and coherent charge density
wave state (UCCDW) at a critical value of the interlayer separation $%
d=d_{c}\left( N\right) .$Working in the generalized random-phase
approximation (GRPA), we have computed the dispersion relations of the
low-energy collective modes of the UCCDW in a range of $d$ where this state
is expected to be the ground state of the 2DEG in the bilayer system. The
UCCDW has two Goldstone modes that are respectively related to the broken
translational symmetry of the stripes and to the broken $U(1)$ symmetry of
the coherent state. In the long-wavelength limit, the dispersion relations
of these modes are consistent with the spin wave dispersion obtained in a
non-collinear ferromagnet with helimagnetic ordering.

\section{Acknowledgments}

The authors thank Profs. D. S\'{e}n\'{e}chal, A.M.-S.
Tremblay, and Drs. Ivana Mihalek
and Luis Brey for helpful discussions.  This work was supported by a grant
from the Natural Sciences and Engineering Research Council of Canada (NSERC) 
and by NSF Grant Nos. DMR-9870681 and DMR-0108451.

{\bf \bigskip }

\newpage

\begin{references}
\bibitem{comcom} The use of long-wavelength commutation relations is only
approximately valid for the states we are studying, which have a pseudospin
density wave. However, the wavelength of this density wave is very large
compared to the magnetic length \cite{shklovskii}, so that the additional
factors in the microscopic operator commutation relations may still be set
to their long-wavelength values without incurring significant error.
\end{references}
\end{document}